\documentclass[aps,prd,reprint,superscriptaddress,nofootinbib]{revtex4-2}
\usepackage[all]{xy}
\usepackage{amsmath,amsthm,amssymb}
\usepackage[dvips]{graphicx}
\usepackage{comment}
\usepackage{array}
\usepackage{bm}
\usepackage{soul}
\allowdisplaybreaks[1]

\usepackage{xcolor}
\usepackage{hyperref}
\usepackage{tcolorbox}
\usepackage{mathrsfs}
\usepackage{float}
\usepackage{orcidlink}
\usepackage{tikz}

\newcommand{\El}{{\sf El}}

\newcommand{\rmin}{r_{\rm min}}

\newcommand{\vf}{\varphi}

\newcommand{\SB}[1]{\left[#1\right]}

\begin{document}

\title{Black hole scattering near the transition to plunge: Self-force and resummation of post-Minkowskian theory}

\def\AEI{Max Planck Institute for Gravitational Physics (Albert Einstein Institute), D-14476 Potsdam, Germany}
\def\Soton{Mathematical Sciences, University of Southampton, 
Southampton, SO17 1BJ, United Kingdom}

\author{Oliver Long \orcidlink{0000-0002-3897-9272}}
\affiliation{\AEI}
\author{Christopher Whittall \orcidlink{0000-0003-2152-6004}}
\affiliation{\Soton}
\author{Leor Barack \orcidlink{0000-0003-4742-9413}}
\affiliation{\Soton}

\date{\today}
\begin{abstract}

Geodesic scattering of a test particle off a Schwarzschild black hole can be parameterized by the speed-at-infinity $v$ and the impact parameter $b$, with a ``separatrix'', $b=b_c(v)$, marking the threshold between scattering and plunge. Near the separatrix, the scattering angle diverges as $\sim\log(b-b_c)$. The self-force correction to the scattering angle (at fixed $v,b$) diverges even faster, like $\sim A_1(v)b_c/(b-b_c)$. Here we numerically calculate the divergence coefficient $A_1(v)$ in a scalar-charge toy model. We then use our knowledge of $A_1(v)$ to inform a resummation of the post-Minkowskian expansion for the scattering angle, and demonstrate that the resummed series agrees remarkably well with numerical self-force results even in the strong-field regime. We propose that a similar resummation technique, applied to a mass particle subject to a gravitational self-force, can significantly enhance the utility and regime of validity of post-Minkowskian calculations for black-hole scattering.
\end{abstract}

\maketitle

\section{Introduction}
The study of the relativistic dynamics in black-hole scattering has been subject of  considerable interest in recent years, motivated in the context of efforts to improve the accuracy and parameter-space reach of theoretical waveform models for gravitational-wave astronomy. Scattering events involving black holes (e.g.~in galactic-center scenarios) are not themselves considered important sources of observable gravitational waves. The idea, rather, is to use information gleaned from scattering analysis to inform precision models of the radiative evolution in {\em bound} binaries of astrophysical interest. The scattering process serves here as a probe of the strong gravitational potential, in much the same way that high-energy particle scattering is used to probe nuclear interaction in particle physics.  The mapping from scattering to bound-orbit dynamics can be achieved either in the framework of Effective One Body (EOB) theory \cite{Damour:2016gwp, Ceresole:2023wxg}, or using certain ``unbound-to-bound'' relations that have been formulated using effective-field theory methods \cite{Goldberger:2004jt,Kalin:2019rwq}. These relations directly map between attributes of scattering and bound orbits, e.g.~between the scattering angle and the periastron advance \cite{Kalin:2019rwq}, or between radiative fluxes \cite{Cho:2021arx}, or even between the emitted waveforms themselves \cite{Adamo:2024oxy}. The scattering setup is fundamentally convenient for analysis, because it admits well-defined `in' and `out' states (with zero binding energy), circumventing some of the coordinate ambiguities that often plague bound-orbit calculations and complicate their interpretation. The study of gravitational scattering has also brought with it the unusual opportunity to apply advanced methods from modern scattering amplitudes theory of particle physics directly to gravity \cite{Bern:2019nnu,Bern:2021dqo, Dlapa:2023hsl}. 

The scattering setup is naturally amenable to a perturbative treatment via the post-Minkowskian (PM) formalism, which a priori restricts the validity of analysis to weak-field scattering at large impact parameters. Much of the progress of the past few years was indeed made in the framework of PM theory \cite{Bern:2021dqo,Bern:2021yeh,Dlapa:2022lmu,Dlapa:2024cje}, or else using PM scattering results to refine EOB models (themselves valid at arbitrary separations) \cite{Damour:2016gwp,Buonanno:2024byg}. Some access into the strong-field regime was made possible using scattering simulations in full Numerical Relativity, enabling important checks on both PM and EOB calculations \cite{Damour:2014afa, Hopper:2022rwo, Rettegno:2023ghr, Albanesi:2024xus}. Another approach to strong-field scattering is via the self-force (SF) formalism, which is based on an expansion in the mass ratio about the limit of geodesic motion, without an expansion in $G$. SF calculations of scattering observables have the potential to enable unique benchmarking of strong-gravity features \cite{Barack:2009ey,Barack:2010ny,Akcay:2012ea}. Thanks to mass-exchange symmetry and the polynomiality of PM expressions in the two masses, SF calculations can also offer a shortcut way of deducing high-order terms in the PM expansion \cite{Damour:2019lcq}. 

Unfortunately, the well-developed SF methods used for modelling bound system of inspiralling black holes \cite{Pound:2021qin,vandeMeent:2017bcc} cannot easily be applied to unbound systems. The technical roadblocks and possible mitigation approaches are surveyed in Refs.\ \cite{Long:2021ufh,Whittall:2023xjp}. As a result, calculations of gravitational scattering in SF theory are still in their infancy. So far actual calculations have been confined to the special example of the marginally-bound trapped orbit \cite{Barack:2019agd}, and to studies involving a double SF-PM expansion \cite{Kosmopoulos:2023bwc, Cheung:2023lnj, Driesse:2024xad}. 

In the meantime, some progress has been made using a scalar-field toy model as a platform for test and development of techniques in preparation for tackling the gravitational problem. In this model (to be reviewed at the end of this introduction) the lighter black hole  is replaced with a pointlike scalar charge that sources a massless Klein-Gordon field, considered as a test field on the fixed geometry of the heavier black hole. One then considers the scattering dynamics under the SF from the scalar field, ignoring the gravitational SF. Within this model, numerical calculations were carried out of the scattering angle in strong-field scenarios, accounting for all scalar-field back-reaction effects, dissipative as well as conservative \cite{Barack:2022pde, Whittall:2023xjp}. A detailed comparison was made with corresponding PM calculations from Amplitudes \cite{Barack:2022pde,Barack:2023oqp}, also illustrating how new high-order PM terms can be numerically determined from SF data \cite{Barack:2023oqp}. The scalar-wave energy absorption by the black hole has also been calculated and shown to agree well with corresponding PM calculations from Amplitudes \cite{Jones:2023ugm,Long:2023Edi}.

Full SF calculations for strong-field scattering are done numerically, since they require solutions of the underlying (linear, or linearized) field equations, which are not known analytically in general.  These calculations can be computationally expensive, although they are not nearly as expensive as full NR simulations; and, unlike most NR simulations, they can return data of very high numerical precision. It is for this reason that SF calculations can be utilized for accurate benchmarking of strong-field aspects of the scattering process. The idea is to use a small set of judiciously chosen bits of SF information in order to inform a ``resummation'' of PM-based analytical formulas, thereby extending their domain of validity into the strong-field regime. This is a computationally cheap(er) alternative to a full SF calculation over the whole parameter space of scattering orbits.

In this paper we use our scalar-field model to illustrate and test this idea. The particular strong-field feature utilized here for that purpose is the singular behavior of the scattering angle at the threshold of transition from scattering to plunging orbits---the so called ``separatrix'' in the parameter space of unbound orbits. We numerically calculate the leading SF correction to that singular behavior (beyond the geodesic-limit expression, known analytically), and use that to inform an accurate resummation formula of the PM expression for the scattering angle. We show how this procedure produces a simple analytical model of the scattering angle that is uniformly accurate on the entire parameter space of scattering orbits. 

A similar resummation strategy was recently adopted by Damour and Rettegno in Ref.\ \cite{Damour:2022ybd} in order to improve the agreement between PM and NR results. In that work use was made of the leading-order, geodesic-limit (logarithmic) form of the separatrix singularity, already producing an impressive improvement. Our work here extends this to include information about the SF term of the singular behavior (albeit in the simpler setting of our scalar-field toy model). This SF term diverges even more strongly than the geodesic-limit term. Then, as we shall see, the resulting improvement is even greater. 

This work required some development of new numerical method, involving a hybridization of our existing time-domain \cite{Barack:2022pde} and frequency-domain \cite{Whittall:2023xjp} codes. The details of this method will be presented in a forthcoming paper \cite{Long_Whittall_inprep}. We will review it here (in Sec.\ \ref{sec:NumMethod}) only briefly. Time-domain and frequency-domain methods perform differently in different areas of the parameter space, and even along the trajectory of a single scattering orbit (for instance, our frequency-domain scheme is extremely accurate near the periapsis but degrades quickly at larger separations). Our new method meshes together data from the two codes to achieve an optimisation of the computational performance. In addition, we extended the reach of our code to greater initial velocities (of up to 0.8$c$), where we discovered that strong radiation beaming necessitated the computation of a very large number of multipolar modes of the scalar field.  There we took advantage of the much superior performance of our frequency-domain scheme at large multipole numbers. 

The structure of the paper is as follows. In Sec.~\ref{sec:ScatGeo+SF} we review scattering orbits in Schwarzschild spacetime, and the calculation of the scattering angle including the leading-order SF effect. In Sec.~\ref{sec:separatrix} we review results concerning the separatrix singularity in the geodesic case, and analyze the form of the singularity when SF is accounted for. We write down formulas that describe the singular behavior in terms of certain integrals of the SF along critical geodesics. In Sec.~\ref{sec:ResumFormula} we introduce our PM resummation formula, which, by design, reproduces the known PM behavior at large values of the impact parameter, as well as the correct singular behavior at the SF-perturbed separatrix. Section \ref{sec:NumMethod} contains a brief review of our numerical method,  and in Sec.~\ref{sec:ResultsA1} we present numerical results for the perturbed separatrix. Section \ref{sec:ResultsResum} then tests our resummation formula against ``exact'' SF data, showing a uniformly good agreement at all values of the impact parameter and for all initial velocities examined. We conclude in Sec.~\ref{sec:Conc} with some general comments and an outlook. 

The rest of this introduction reviews the scalar-field model used in this work. Throughout this work we use geometrized units, with $G=c=1$. 

\subsection{Scalar-field toy model}

We consider a pointlike particle carrying a scalar charge $q$ and mass $\mu$ in a scattering orbit around a Schwarzschild black hole of mass $M\gg\mu$. The particle sources a scalar field $\Phi$, assumed to be governed by the massless, minimally coupled Klein-Gordon equation
\begin{equation}\label{eq: KG with source}
	g^{\mu \nu} \nabla_\mu \nabla_\nu \Phi = -4 \pi q \int \frac{\delta^4 ( x^{\alpha} - x^{\alpha}_p (\tau))}{\sqrt{-g}} d\tau,
\end{equation}
where $g^{\mu \nu}$ is the inverse Schwarzschild metric and $\nabla_\mu$ is the covariant derivative compatible with it. On the right-hand side, $g$ is the metric determinant and  $x^{\alpha}_p(\tau)$ describes the particle's worldline, parameterized in terms of proper time $\tau$. The field $\Phi$ is assumed to satisfy the usual retarded boundary conditions at null infinity and on the event horizon. 

In the limit where both $\mu\to 0$ and $q\to 0$, the particle follows a timelike geodesic of the background Schwarzschild metric $g_{\alpha\beta}$, satisfying $u^\beta \nabla_\beta ( \mu u^\alpha)= 0 $, where $u^\alpha:= d x_p^\alpha/d\tau$ is the tangent four-velocity. For a finite $q$, the particle experiences a SF $\propto q^2$ due to back reaction from $\Phi$. This accelerates the particle's worldline away from geodesic motion. The magnitude of self-acceleration is controlled by the dimensionless parameter
\begin{equation}\label{eq:eps}
\epsilon: = \frac{q^2}{\mu M}.
\end{equation} 
We assume $\epsilon\ll 1$, so that the worldline is only slightly perturbed off the original geodesic, by an amount $\propto\epsilon$. The particle's equation of motion now reads
\begin{equation}\label{eq:cov_evolution}
	\mu u^\beta \nabla_\beta u^\alpha = q (g^{\alpha\beta}+u^\alpha u^\beta )\nabla_\beta \Phi^R =: F^\alpha_{\rm self},
\end{equation}
where $\Phi^R$ is the Detweiler-Whiting regular piece of $\Phi$ (`R field') at the position of the particle \cite{Detweiler:2002mi}. The SF term on the right-hand side here accounts for both conservative and dissipative effects of the scalar-field back-reaction.  

In this work we completely neglect the {\em gravitational} back-reaction on the particle's motion, as well as any back-reaction from $\Phi$ on the background spacetime itself: the field $\Phi$ is treated here as a test field. We also ignore the small change in $\mu$ caused by the component of $\nabla_\beta \Phi^R$ tangent to $u^\alpha$ (interpreted as an exchange of energy between the particle and the scalar field). 

\section{Scattering geodesics and their self-force perturbation}
\label{sec:ScatGeo+SF}

We start by reviewing relevant results from the theory of scattering geodesics and their SF perturbation in Schwarzschild spacetime. We use Schwarzschild coordinates $(t,r,\theta,\varphi)$ attached to the black hole of mass $M$, and without loss of generality take the scattering orbit to lie in the equatorial plane, $\theta = \pi/2$. From symmetry, the orbit remains in the equatorial plane even under the SF effect. Note that the system's center of mass is fixed at the origin of our Schwarzschild coordinates, since we neglect the gravitational effect of $\mu$.  

\subsection{Geodesic limit}

In the limit $\epsilon\to 0$, $x_p(\tau)$ is a timelike geodesic with conserved energy and angular momentum given (per $\mu$) by
\begin{align}
	 E &:= (1-2M/r_p)\,  \dot{t}_p\label{eq:tdot},\\
	 L &:= r_p^2 \, \dot{\varphi}_p\label{eq:phidot},
\end{align}
where an overdot denotes $d/d\tau$. We are interested in a scattering scenario, where  $r_p\to\infty$ as $\tau\to\pm\infty$. This requires $E > 1$ and $L>L_{\rm c}(E)$, where $L_{\rm c}(E)$ describes the separatrix between scattering and captured geodesics, to be analyzed in more detail in Sec.~\ref{sec:separatrix}.  The pair $(E,L)$ can be used to parametrize the family of timelike scattering geodesics. Alternatively, we can use the pair $(v,b)$, where $v$ is the magnitude of the 3-velocity at infinity,
\begin{align}\label{v}
	v &:= \lim_{\tau\to-\infty}\sqrt{\left(\dot r_p^2 + r_p^2 \dot\varphi_p^2\right)/\dot t_p^2} = \frac{\sqrt{E^2-1}}{E},
\end{align}
and $b$ is the impact parameter,
\begin{align}
	b := \lim_{\tau\to-\infty} r_p(\tau)\sin|\varphi_p(\tau)-\varphi_p(-\infty)| = \frac{L}{vE}.\label{eq:bdef}
\end{align}	
The orbit is then a scattering geodesic provided
\begin{align}
	b > b_{\rm c}(v) := \frac{L_{\rm c}(E)}{vE},
\end{align}
where $E=(1-v^2)^{-1/2}$.

We let $\varphi_{\rm in}:=\varphi_p(\tau\to -\infty)$ and $\varphi_{\rm out}:=\varphi_p(\tau\to +\infty)$. The \textit{scattering angle} is then defined to be
\begin{align}
	\chi^{\rm 0SF} := \vf_{\rm out} - \vf_{\rm in} - \pi,
\end{align}
where hereafter we use the label `0SF' to denote geodesic-limit values. The geodesic-limit scattering angle is given explicitly by (see, e.g., \cite{Barack:2022pde})
\begin{align}
	\chi^{\rm 0SF} = 2k\sqrt{p/e}\, \El_1\left(\psi, -k^2\right) - \pi. \label{eq:geodesic_scatter_angle}
\end{align}
Here $p$ and $e$ are the (unique) solutions of 
\begin{align}
	E^2 = \frac{(p-2)^2 - 4e^2}{p(p-3-e^2)}, \quad\quad L^2 = \frac{p^2M^2}{p-3-e^2} \label{eq:ELepRelation}
\end{align}
satisfying $e>1$ and $p>6+2e$, and we have introduced $\psi := \frac{1}{2}\arccos{(-1/e)}$,
$k:= \sqrt{4e/(p-6-2e)}$, and the incomplete elliptic integral of the first kind,
\begin{equation}
\El_1(\psi, z) = \displaystyle\int_0^{\psi} \frac{d\theta}{\sqrt{1-z\sin^2 \theta}}\, .
\end{equation}
The pair $(p,e)$ forms a `geometrical' parametrization of geodesic scattering orbits, with $p$ interpreted as semilatus rectum (divided by $M$) and $e$ as eccentricity. In the $p$--$e$ plane, the separatrix takes the very simple form $p_c(e)=6+2e$, with $p>p_c(e)$ for scattering orbits. 

\subsection{1SF correction}

When the SF term is included on the right-hand side of the equation of motion (\ref{eq:cov_evolution}), the solution $x_p(\tau)$ is no longer a geodesic of the Schwarzschild background but a slightly accelerated worldline. For fixed values of $v$ and $b$, the scattering angle picks up an $O(\epsilon)$ correction with respect to its geodesic value. We write the perturbed angle in the form
\begin{equation}
\chi(v,b) = \chi^{\rm 0SF}(v,b) + \epsilon\,\chi^{\rm 1SF}(v,b),
\end{equation}
where the split between the 0SF (geodesic) term and the 1SF (self-force) term is defined at fixed $v,b$.  An expression for $\chi^{\rm 1SF}$ in terms of an integral of the self-force along the orbit was derived in Ref.~\cite{Barack:2022pde}:
\begin{align}\
    \chi^{\rm 1SF} = \int_{-\infty}^{\infty} \SB{\mathcal{G}_E(\tau)\tilde F^{\rm self}_t(\tau) - \mathcal{G}_L(\tau)\tilde F^{\rm self}_{\varphi}(\tau)}d\tau\label{eq:1SF_scatter_angle},
\end{align}
where $\tilde F^{\rm self}_\alpha:=(M/q^2) F^{\rm self}_\alpha$, and, within our approximation, it suffices to evaluate the SF components along the background geodesic. The functions $\mathcal{G}_{E}(\tau)$ and $\mathcal{G}_{L}(\tau)$ are also evaluated along the background geodesic, and depend on its parameters; these functions are given explicitly (in terms of $p,e$) in Sec.\ IV.A of Ref.~\cite{Barack:2022pde}. 

In practice, and also in aiding comparison with PM results, it is convenient to split the SF into its conservative and dissipative pieces, $F^{\rm self}_{\alpha}=F^{\rm cons}_{\alpha}+F^{\rm diss}_{\alpha}$, and correspondingly write $\chi^{\rm 1SF}$ as a sum of conservative and dissipative contributions, $\chi^{\rm 1SF}=\chi^{\rm cons}+\chi^{\rm diss}$. The separate pieces are obtained via \cite{Barack:2022pde}
\begin{align}
    \chi^{\rm cons} = \int_{0}^{\infty} \SB{\mathcal{G}^{\rm cons}_E(\tau)\tilde F^{\rm cons}_t(\tau) - \mathcal{G}^{\rm cons}_L(\tau)\tilde F^{\rm cons}_{\varphi}(\tau)}d\tau, \label{eq:1SF_scatter_angle_cons}
\end{align}
\pagebreak
and
\begin{align}
    \chi^{\rm diss} &= \int_{0}^{\infty} \SB{\beta_E\, \tilde F^{\rm diss}_t(\tau) - \beta_L\, \tilde F^{\rm diss}_{\varphi}(\tau) } d\tau
    \nonumber\\
    &=
    -\frac{1}{2}\left(\beta_E E_{\rm rad} + \beta_L L_{\rm rad}\right),
    \label{eq:1SF_scatter_angle_diss}
\end{align}
where $\tau=0$ corresponds to periastron passage. The functions $\mathcal{G}^{\rm cons}_{E,L}(\tau)$ and the coefficients $\beta_{E,L}$ are given explicitly in (respectively) Secs.\ IV.B and V.C of \cite{Barack:2022pde}, in terms of the orbital parameters $p,e$. The second line of (\ref{eq:1SF_scatter_angle_diss}) expresses $\chi^{\rm diss}$ in terms of the total energy $E_{\rm rad}$ and angular momentum $L_{\rm rad}$ (per $q^2/M$) radiated in scalar-field waves during the entire scattering process. 

Since a calculation of $\tilde F^{\rm self}_\alpha$ involves solving the field equation (\ref{eq: KG with source}), which can only be done numerically in general, the value of $\chi^{\rm 1SF}$ (and of either of its separate pieces $\chi^{\rm cons}$ and $\chi^{\rm diss}$) can only be obtained numerically, in general, for each specific values of $v,b$. In practice, $v,b$ need first be converted to $p,e$, which is readily done using Eqs.\ (\ref{v}), (\ref{eq:bdef}) and (\ref{eq:ELepRelation}).

\subsection{Post-Minkowskian expansion}

Approximate analytical solutions for $\chi^{\rm 1SF}$ can be obtained order by order in a PM expansion. The expansion takes the form
\begin{equation}\label{PM series}
\chi^{\rm 1SF}= \sum_{k=2}^\infty \chi_k^{{\rm 1SF}}(v) \left(\frac{G M}{b}\right)^n,
\end{equation}
where we have temporarily reinstated $G$ for clarity. A similar expansion holds for $\chi^{\rm cons}$ and $\chi^{\rm diss}$, with expansion coefficients to be denoted $\chi_k^{{\rm cons}}$ and $\chi_k^{{\rm diss}}$, respectively. The PM coefficients known so far are
\begin{widetext}
\begin{eqnarray}
\chi_2^{\rm cons} &=& -\frac{\pi}{4}, \label{chi2cons}\\
\chi_3^{\rm cons} &=& -\frac{4E \left(3-v^2\right)}{3 v^2}, \label{chi3cons}
\end{eqnarray}
\begin{eqnarray}
\chi_4^{\rm cons} &=& \frac{\pi}{32 v^5 E^4} \Bigg[ - 6 (95 E+82) v \: \El_1\left(\frac{\pi}{2},\frac{E-1}{E+1}\right)^2 + 6 (E (100 E+177)+79) v \: \El_1\left(\frac{\pi}{2},\frac{E-1}{E+1}\right) \El_2 \left(\frac{\pi}{2},\frac{E-1}{E+1}\right)\nonumber\\
&&\hspace{1.5cm}- \: 3 (E+1) \left(100 E^2+79\right) v \: \El_2 \left(\frac{\pi}{2},\frac{E-1}{E+1}\right)^2 + 9 E^6 v \left(1-3 v^2\right)^2 \text{arccosh}^2(E) \nonumber\\
&&\hspace{1.5cm}+\: E^6 \left(1-3 v^2\right) \left(36 v^4 \log \left(\frac{E v}{2}\right)-29 \left(2-v^2\right) v^2-16\right)\text{arccosh}(E) +48 E^4 v^5 \log \left(b/M\right) \nonumber\\
&&\hspace{1.5cm}+\:2 E^6 v^3 \left((38-24 E) v^4+(24 E-58) v^2-16\right) \log \left(\frac{E v}{2}\right) -36 E^6 v^7 \log ^2\left(\frac{1+E}{2}\right) \nonumber\\
&&\hspace{1.5cm}+\: 6 E^6 v^3 \left((8 E-27) v^4+12 v^4 \log \left(\frac{E v}{2}\right)-8 (E-4) v^2-8\right)\log \left(\frac{1+E}{2}\right) \nonumber \\
&&\hspace{1.5cm}\:-v \left[18 E^6+252 E^5-216 E^3+463 E^2-348 E+E^4 \left(12 v^4+8 v^2-223\right)+110\right] \Bigg], \label{chi4cons} \label{chi4cons}
\end{eqnarray}
and
\begin{eqnarray}
\chi_2^{\rm diss} &=& 0 ,\label{chi2diss}\\
\chi_3^{\rm diss} &=& \frac{2 E}{3} \frac{\left(1+v^2\right)^2}{v^3} ,\label{chi3diss}
\end{eqnarray}
\begin{eqnarray}
\chi_4^{\rm diss} &=&
	\frac{\pi  E}{8 v}
    \Bigg[
   \frac{3 E \left(1-3 v^2\right) \left(1+5 v^2\right)}{2 v^3}{\rm arccosh}(E)
    +3 E \left(1+5 v^2\right) \log \left(\frac{1+E}{2}\right)
    \nonumber \\
    &&\qquad\quad+\:\frac{24 E+(61 E+18) v^6+2 (75-52 E) v^4+(19 E+84) v^2}{6 v^4}
    \Bigg], \label{chi4diss}
\end{eqnarray}
\end{widetext}
where $\El_2$ is the incomplete elliptic integral of the second kind: 
\begin{equation}
\El_2(\psi, z) = \displaystyle\int_0^{\psi} \sqrt{1-z\sin^2 \theta} \: d\theta.
\end{equation}
The leading PM term $\chi_2^{\rm cons}$ was first obtained in \cite{Gralla:2021qaf}, and the leading term $\chi_3^{\rm diss}$ was first obtained in \cite{Barack:2022pde} (using results from \cite{Gralla:2021qaf}). The terms $\chi_3^{\rm cons}$, $\chi_4^{\rm cons}$ and $\chi_4^{\rm diss}$ were derived in \cite{Barack:2023oqp} using Amplitude methods, except for certain pieces of the polynomial expression in the last line of Eq.\ (\ref{chi4cons}) for $\chi_4^{\rm cons}$, associated with undetermined Wilson coefficients. These have more recently been determined through two independent methods: Matching of amplitude calculations in effective-field theory and black hole perturbation theory \cite{Ivanov:2024sds}; and a double PM/post-Newtonian expansion of the self-force \cite{Bini:2024icd}.

We will use notation whereby $\chi^{\rm 1SF}_{n{\rm PM}}$ ($n\geq 2$) represents the truncated sum $\sum_{k=2}^n$ in Eq.\ (\ref{PM series}), and we will analogously have $\chi^{\rm cons}_{n{\rm PM}}$ and $\chi^{\rm diss}_{n{\rm PM}}$. To denote the $n$-th PM order truncation of the full scattering angle, $\chi^{\rm 0SF}+\epsilon \chi^{\rm 1SF}$, we will use $\chi_{n{\rm PM}}$. 

\begin{figure}[htb]
\centering
\includegraphics[width=\linewidth]{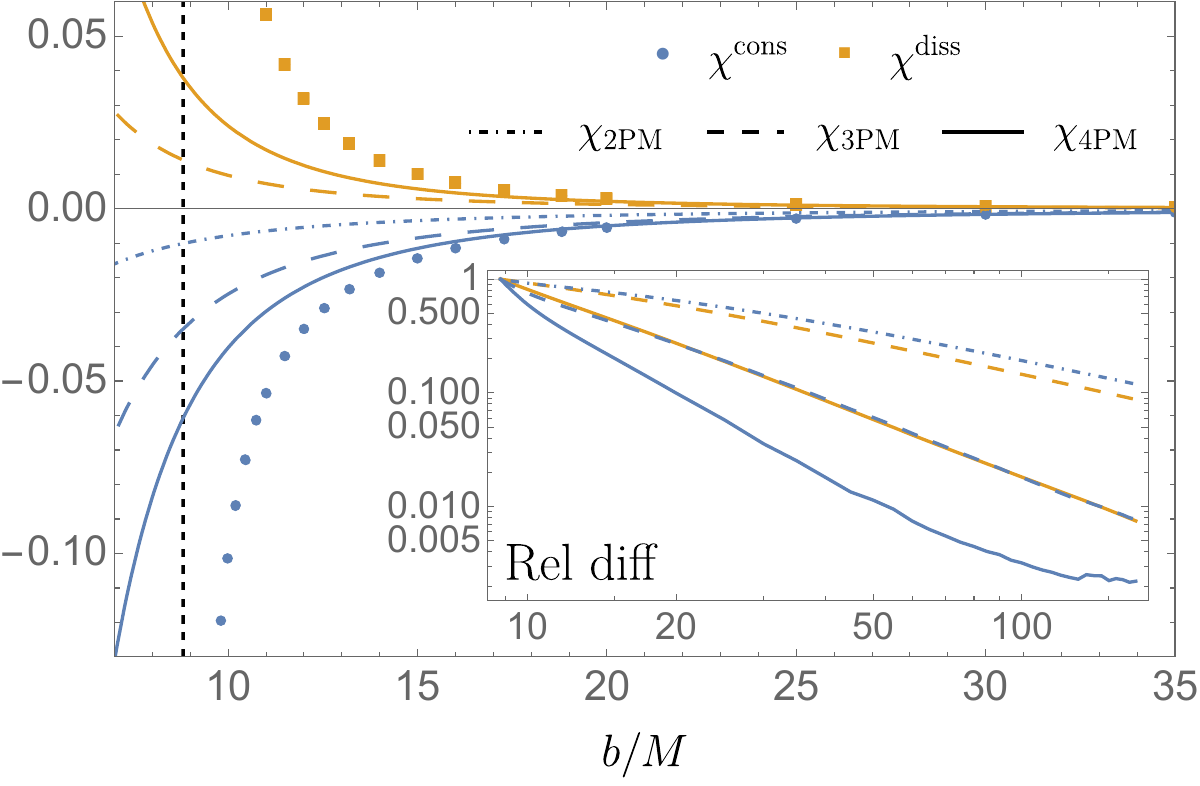}
\caption{Comparison of successive PM approximations with ``exact'' numerical SF values for the conservative (blue) and dissipative (orange) contributions to the 1SF scattering angle at $v=0.5$. The inset shows the relative difference between the analytical PM approximation and the numerical SF values (interpolated over $b$). The PM expressions accurately describe the weak-field behavior, with 4PM conservative results being within the numerical error of the data for $b\gtrsim 100M$. The PM expressions break down when approaching the separatrix, here at $b\simeq 8.807M$ (vertical dashed line), where $\chi^{\rm 1SF}$ diverges. The numerical values were generated using the time-domain code detailed in Sec.\ \ref{sec:TD}. The numerical errors are too small to be resolved on the scale of the main plot. 
}
\label{Fig:PM_vs_SF}
\end{figure}

Figure \ref{Fig:PM_vs_SF} shows the sequence of PM approximations $\chi^{\rm 1SF}_{n{\rm PM}}$ for $n=2,3,4$ as functions of $b$ at fixed $v=0.5$. For reference, a sample of ``exact'' SF results is also shown, obtained using the (time-domain) numerical method reviewed in Sec.\ \ref{sec:TD}. As expected, the agreement is increasingly better at larger $b$ and greater $n$, but the PM expressions fail to capture the true behavior at small $b$ and especially near the separatrix. The goal of this paper is to demonstrate how, utilising information about the singular form of the SF near the separatrix, one can resum the PM series to the effect of making it a uniformly good approximation at all $b$.  

\section{The separatrix singularity and its self-force perturbation}\label{sec:separatrix}

\subsection{Geodesic limit}

The separatrix is a 1-dimensional curve in the 2-dimensional parameter space, which separates between scattering and plunging geodesics. It is given by $b=b_c(v)$, where
\begin{equation}\label{bc}
b_c(v) = \frac{M}{\sqrt{2}\,v^2}\sqrt{8v^4+\beta-1 +4v^2(2\beta+5)},
\end{equation} 
with 
\begin{equation}
\beta:=\sqrt{1+8v^2}.
\end{equation}
The function $b_c(v)$ is monotonically decreasing in $v$, with
\begin{align}
\lim_{v\to 0} b_c(v)=&\infty,\\
\lim_{v\to 1}b_c(v)=&3\sqrt{3}\,M\simeq 5.196\,M.
\end{align}
The latter value bounds from below the overall range of $b$ values possible for scattering geodesics. We refer to members of the 1-parameter family of geodesics that form the separatrix, i.e., ones with $b=b_c(v)$, as ``critical geodesics''. Each critical geodesic is made up of two disjoint branches: an ``inbound'' branch starting at infinity and ending with an infinite circular whirl at periastron distance, and its time-reversed ``outbound'' branch.   

Near the separatrix, the geodesic scattering angle $\chi^{\rm 0SF}$ diverges logarithmically:
\begin{equation}\label{chi0_bv}
\chi^{\rm 0SF} = A_0(v)\log\left(\frac{\delta b}{b_c(v)}\right) + {\rm const}(v) + \cdots
\end{equation}
where $\delta b:= b-b_c(v)$, and 
\begin{equation}\label{A0}
A_0(v) = -\left(1-\frac{12M^2(1-v^2)}{v^2 b_c(v)^2}\right)^{-1/4}. 
\end{equation}
A derivation of this asymptotic expression is described in Appendix \ref{App}. 

\subsection{1SF correction}

We have assumed that the SF perturbed the orbit by only a small amount, such that its leading-order correction to the scattering angle can be obtained by integrating along the background geodesic. This assumption must ultimately fail as we get near enough the separatrix, where the background geodesic gets trapped in an eternal whirl at periastron distance. Due to radiative losses, the actual SF-perturbed orbit does not exercise an infinite whirl, but instead it either falls into the black hole or scatters off back to infinity after a finite amount of whirl time.

However, we can still have a well-defined notion of SF correction to the scattering angle even near the separatrix, by continuing to integrate along the background geodesic. The so-defined SF correction  $\chi^{\rm 1SF}$ need not be ``small'' compared to $\chi^{\rm 0SF}$ near the separatrix. In fact, the numerical results of Ref.\ \cite{Barack:2022pde} (cf.\ Figure 9 therein) suggest a stronger divergence at 1SF order, of the form
\begin{equation}\label{chi1SF_asy}
\chi^{\rm 1 SF}\sim A_1(v)\left(\frac{b_c(v)}{\delta b}\right).
\end{equation}
In principle, if desired, we can still ensure $\epsilon\chi^{\rm 1SF}\ll \chi^{\rm 0SF}$ simply by taking $\epsilon\to 0$ sufficiently fast as we take $\delta b\to 0$. 

We now derive the asymptotic relation (\ref{chi1SF_asy}) analytically, and obtain an expression for $A_1(v)$ in terms of SF integrals along critical geodesics. 

Considering first the dissipative piece, our starting point is the general formula for $\chi^{\rm diss}$ in the first line of Eq.\ (\ref{eq:1SF_scatter_angle_diss}). Our goal is to approximate this expression at small $\delta b = b-b_c(v)$ for an arbitrary $v$. To this end, we substitute $p=p_c(e)+\delta p=6+2e+\delta p$ in $\beta_{E,L}$, expand to leading order in $\delta p$ (at fixed $e$), and then substitute for $\delta p$ in terms of $\delta b$ using the leading-order expression (\ref{deltab_deltap}) from the appendix. This procedure gives, at leading order in $\delta b$,
\begin{equation}\label{deltavarphi_diss_critical}
\chi^{\rm diss}\sim \frac{1}{\delta b}
\int_{-\infty}^{\infty} \left(c_E \tilde F_t^{\rm diss}
+c_L  \tilde F_\varphi^{\rm diss}\right)d\tau,
\end{equation}
with 
\begin{align}
c_E &= -\frac{2(3-e)^{1/2}(3+e)^{5/2}M}{(e+1)^2 \sqrt{e(e-1)}},\label{eq:cE}
\\
c_L &= -\frac{(3+e)(3-e)^{1/2}}{\sqrt{2e(e^2-1)}}.\label{eq:cL}
\end{align}
Here the integrand is evaluated along the outbound branch of the critical geodesic with $b=b_c(v)$, whose periastron is at $\tau\to -\infty$. We have thus reproduced Eq.\ (\ref{chi1SF_asy}), with
\begin{equation}\label{A1_diss}
A_1^{\rm diss}(v) = 
\frac{1}{b_c(v)}\int_{-\infty}^{\infty} \left(c_E \tilde F_t^{\rm diss}
+c_L  \tilde F_\varphi^{\rm diss}\right)d\tau .
\end{equation}
Note that for critical orbits we have $1<e<3$, so expressions like those in (\ref{eq:cE}) and (\ref{eq:cL}) make sense. One should resist temptation to write the integral here as $-c_E E_{\rm rad}+ c_L L_{\rm rad}$ [recalling Eq.\ (\ref{eq:1SF_scatter_angle_diss})], since both $E_{\rm rad}$ and $L_{\rm rad}$ diverge for a critical geodesic, due to the infinite whirl. The integral in Eq.\ (\ref{A1_diss}), however, is well defined and finite. We discuss the convergence of this integral further below, in Sec.\ \ref{sec:convergence}.

A similar procedure is applied to the conservative piece. Starting with Eq.\ (\ref{eq:1SF_scatter_angle_cons}), we expand the functions $\mathcal{G}^{\rm cons}_E(\tau)$ and $\mathcal{G}^{\rm cons}_L(\tau)$ (given explicitly in \cite{Barack:2022pde} in terms of $p,e$) in $\delta p$ about 
$p_c(e)+\delta p=6+2e$ at fixed $e$ and $\tau$, and then substitute for $\delta p$ in terms of $\delta b$ using the leading-order expression (\ref{deltab_deltap}) from the appendix. We find that the leading-order term in $\delta b$ is $\tau$-independent. In fact, the calculation yields an expression of a very similar form to that for $\chi^{\rm diss}$:
\begin{equation}\label{deltavarphi_cons_critical}
\chi^{\rm cons}\sim -\frac{1}{\delta b}
\int_{-\infty}^{\infty} \left(c_E \tilde F_t^{\rm cons}
+c_L  \tilde F_\varphi^{\rm cons}\right)d\tau,
\end{equation}
with the same coefficients $c_E$ and $c_L$ as in Eq.\ (\ref{deltavarphi_diss_critical}). Thus
\begin{equation}\label{A1_cons}
A_1^{\rm cons}(v) = -\frac{1}{b_c(v)}
\int_{-\infty}^{\infty} \left(c_E \tilde F_t^{\rm cons}
+c_L  \tilde F_\varphi^{\rm cons}\right)d\tau .
\end{equation}
Again, the integration is done along the outbound branch of the critical geodesic with velocity $v$. Its convergence is discussed in Sec.\ \ref{sec:convergence}.

The complete coefficient $A_1(v)=A_1^{\rm diss}(v) + A_1^{\rm cons}(v)$ 
can be written in terms of the full SF $\tilde F_\alpha^{\rm self}$, by recalling that, if we think of the SF as a function of $r$ and $\dot{r}$ on the critical geodesic, then we have the symmetries $\tilde F_t^{\rm diss}(r,\dot{r})=\tilde F_t^{\rm diss}(r,-\dot{r})$ and $\tilde F_t^{\rm cons}(r,\dot{r})=-\tilde F_t^{\rm cons}(r,-\dot{r})$ [cf.\ Eq.\ (42) of \cite{Barack:2022pde}]. This means that the value of $\tilde F_t^{\rm diss}-\tilde F_t^{\rm cons}$ at a given point on the outbound branch is equal to  $\tilde F_t^{\rm diss}+\tilde F_t^{\rm cons}= \tilde F_t^{\rm self}$ at a conjugate point with the same $r$ value on the inbound branch, and similarly for the $\varphi$ components.  Thus we obtain
\begin{align}\label{A1}
A_1(v) = \frac{1}{b_c(v)}
\int_{-\infty}^{\infty} \big(c_E \tilde F_t^{\rm self} +c_L  \tilde F^{\rm self}_\varphi\big) d\tau\, ,
\end{align}
where the integration is now performed along the {\em inbound} leg of the orbit, starting at infinity ($\tau\to -\infty$) and ending at the whirl radius ($\tau\to +\infty$). In Eq.\ (\ref{A1}), as in (\ref{A1_diss}), the individual integrals over the $c_E$ and $c_L$ terms do not exist, but the integral of their sum does. This we explain in what follows. 

\subsubsection{Convergence of integrals}\label{sec:convergence}

At large radius, $F_t$ and $F_{\varphi}/r_p$ decay as $r_p^{-3}$ \cite{Barack:2022pde}, so the integrals in Eqs.\ (\ref{A1_diss}), (\ref{A1_cons}) converge well at their upper limits, and the one in (\ref{A1}) converges well at its lower limit. As mentioned, the convergence in the opposite limit, where the critical geodesic executes an infinite whirl, is more subtle, and requires some analysis. 

Start with the conservative case. For a nearly circular orbit we have $F_\alpha^{\rm cons}\propto \dot{r}_p$ for $\alpha=t,\varphi$ \cite{Barack:2010tm}, so (\ref{A1_cons}) can be written as 
\begin{equation}\label{A1_cons_r}
A_1^{\rm cons}(v) = -\frac{1}{b_c(v)}
\int_{r_{\rm min}}^{\infty} \left(c_E \hat F_t^{\rm cons}
+c_L  \hat F_\varphi^{\rm cons}\right)dr ,
\end{equation}
where $r_{\rm min}$ is the whirl (periastron) radius, and $\hat F^{\rm cons}_{\alpha}:=\tilde F^{\rm cons}_{\alpha}/\dot{r}_p$ has a finite $r\to r_{\rm min}$ limit. In fact, each of the components are bounded everywhere in the integration domain, and fall off as $\hat F^{\rm cons}_{t}\sim1/r^3$ and $\hat F^{\rm cons}_{\varphi}\sim1/r^2$ at large $r$, so the integral in (\ref{A1_cons_r}) exists. 

The dissipative case is more delicate. The components $F_t^{\rm diss}$ and $F_\varphi^{\rm diss}$ do not vanish on the whirl radius, so the integrals of the corresponding terms in Eq.\ (\ref{A1_diss}) do not separately converge, and the same holds true also for the two separate full SF integrals in Eq.\ (\ref{A1}) (since the integrals of the conservative pieces do converge). However, the integral of the sum of $c_E$ and $c_L$ terms does converge, in both (\ref{A1_diss}) and (\ref{A1}). To see this, note 
\begin{equation}
\frac{c_L}{c_E}  = \frac{1}{M}\left(\frac{1+e}{6+2e}\right)^{3/2}=\sqrt{\frac{M}{r_{\rm min}^3}}=\Omega, 
\end{equation}
the angular frequency ($=u^\varphi/u^t$) of the whirl. Near the whirl radius we have $u^\varphi/u^t=\Omega +O(\dot{r}_p)$, so, using $u^\alpha \tilde F^{\rm self}_\alpha=0$, it follows that, during the whirl,
\begin{align}\label{integrand}
c_E \tilde F_t^{\rm self}+c_L  \tilde F_\varphi^{\rm self}
&= c_E (\tilde F_t^{\rm self}+\Omega  \tilde F_\varphi^{\rm self}) \nonumber\\
&= - c_E (\dot{r}_p/\dot t_p)\tilde F_r^{\rm self} +O(\dot{r}_p).
\end{align}
Thus the integrand in Eq.\ (\ref{A1}) is bounded everywhere (falling off as $r_p^{-3}$ at infinity), and we conclude that the integral converges. Since the integral of the conservative piece converges, we can conclude that, in Eq.\ (\ref{A1_diss}), the integral of the dissipative piece alone also converges. 

Nonetheless, care must be taken when numerically evaluating the integrands in Eqs.\ (\ref{A1_diss}) and (\ref{A1}) near the whirl radius, due the large degree of cancellation expected between the $c_E$ and $c_L$ terms. 

\section{PM Resummation formula}
\label{sec:ResumFormula}

We now suggest a way of using knowledge of the SF singularity coefficient $A_1(v)$ to inform a PM resummation with an improved performance at small $b$.

Consider the function
\begin{align} \vspace{-2mm} \label{eq:ResumFormula}
\Delta\chi(b,v) = A_0(v) \Bigg[ &\log\left(1- \frac{b_c(v)[1-\epsilon A_1(v)/A_0(v)]}{b}\right)
\nonumber\\
&+ \sum_{k=1}^{4} \frac{1}{k}\left(\frac{b_c(v)[1-\epsilon A_1(v)/A_0(v)]}{b}\right)^k \Bigg].
\end{align}
It has the following properties. 
(i) It is of 5PM order: $\Delta\chi = O(b^{-5})$ at $b\gg b_c(v)$; the term in the second line is designed to cancel out all lower-order PM terms of the expression in the first line. 
(ii) In the geodesic limit, $\epsilon\to 0$, $\Delta\chi$ has the same logarithmic divergence as $\chi^{\rm 0SF}$ near the separatrix [compare with Eq.\ (\ref{chi0_bv})], so that $\lim_{b\to b_c(v)}\left(\chi^{\rm 0SF}-\Delta\chi^{\rm 0SF}\right)$ is finite. Here we have introduced the expansion $\Delta\chi= \Delta\chi^{\rm 0SF}+\epsilon \Delta\chi^{\rm 1SF}+O(\epsilon^2)$, as always defined with fixed $v,b$. 
(iii) At 1SF order [$O(\epsilon)$], $\Delta\chi$ exhibits the same $\sim 1/\delta b$ divergence as $\chi^{\rm 1SF}$, with the same coefficient $A_1(v)$:
\begin{equation}
\Delta\chi^{\rm 1SF} \sim A_1(v)\left(\frac{b_c(v)}{\delta b}\right).
\end{equation}
Note that the quantity $-\epsilon \: b_c(A_1/A_0)$ in Eq.\ (\ref{eq:ResumFormula}) may be interpreted as the self-force correction to the critical impact parameter $b_c(v)$ (at fixed $v$).

We then introduce the {\it resummed} scattering angle
\begin{equation}
\tilde\chi(b,v) := \chi_{4 {\rm PM}}(b,v) + \Delta\chi(b,v). \label{eq:ResumFormula2}
\end{equation}
It has the following properties.
(1) By virtue of above property (i), the 4PM truncation of $\tilde\chi$ is identical to $\chi_{\rm 4PM}$, i.e.~$\tilde \chi_{4 {\rm PM}} = \chi_{4 {\rm PM}}$.
(2) By virtue of above property (ii), combined with the fact that $\chi_{4 {\rm PM}}$ is regular at the separatrix, we have that, in the geodesic limit, $\tilde\chi$ has the same logarithmic divergence as $\chi^{\rm 0SF}$ near the separatrix.
(3) By virtue of above property (iii), $\tilde\chi^{\rm 1SF}$ has the same $\sim b_c/\delta b$ divergence as $\chi^{\rm 1SF}$ near the separatrix, with the same coefficient $A_1(v)$.
Thus $\tilde\chi$ reproduces the asymptotic behavior of $\chi$ at $b\to\infty$ through 4PM order, and its asymptotic behavior at $b\to b_c(v)$ through 1SF order.

\begin{figure}[htb]
\centering
\includegraphics[width=\linewidth]{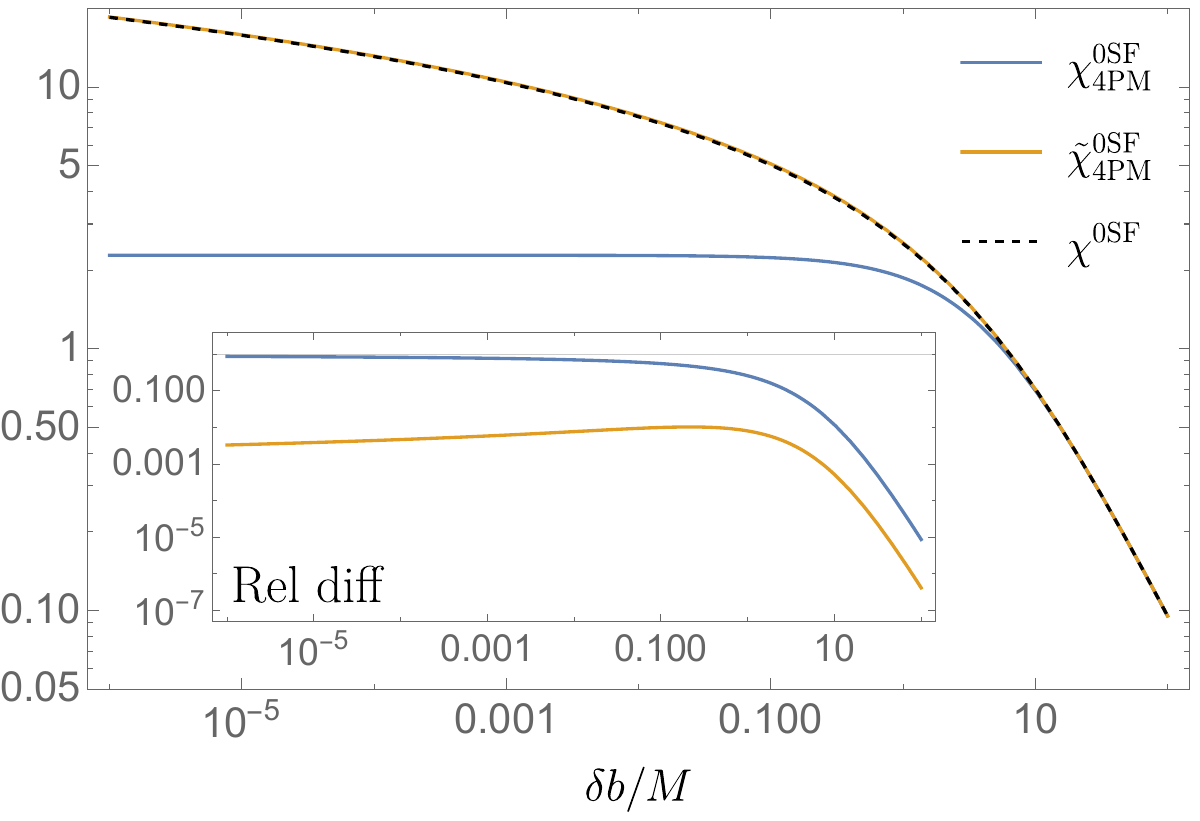}
\caption{Comparison of plain and resummed 4PM expressions for the geodesic scattering angle at $v=0.5$. The exact geodesic expression $\chi^{\rm 0SF}$ from Eq.\ (\ref{eq:geodesic_scatter_angle}) is shown for reference in black dashed line. The plain 4PM expression (blue) fails to capture the logarithmic divergence near the separatrix. In the resummed version (orange) we force the correct singular behavior, and this dramatically improves the performance of the PM model everywhere. The inset shows the relative difference between the full expression and the plain 4PM expression (blue) and between it and the resummed one (orange).
}
\label{Fig:GeoResum}
\end{figure}

We can immediately test the utility of our $\tilde\chi$ in the geodesic limit, without any SF calculation. An illustration is presented in Fig.\ \ref{Fig:GeoResum} for $v=0.5$. The plot compares the full $\chi^{\rm 0SF}$ expression [Eq.\ (\ref{eq:geodesic_scatter_angle})] with its 4PM truncation $\chi^{\rm 0SF}_{\rm 4PM}$, and with the resummed version $\tilde\chi^{\rm 0PM}$ of its 4PM truncation. Evidently, by forcing the PM expression to emulate the correct logarithmic divergence at the separatrix, we dramatically improve its faithfulness at all $b$. A striking feature is that the resummation appears to improve the faithfulness of the PM expression even in the PM domain at large $b$. 

Of course, there is hardly a need to introduce resummation in the geodesic case, where a simple exact expression for $\chi$ is at hand.  The method becomes advantageous at 1SF, where no analytical expression exists and numerical calculations are expensive. The only numerical input necessary for $\tilde\chi$ is the value of the singularity coefficient $A_1(v)$. In principle, this requires evaluation of the SF only for the 1-parameter family of critical geodesics, which should be much cheaper than a full coverage of the 2-dimensional parameter space. 

In what follows we describe our numerical method for calculating $A_1(v)$, present the values obtained (with an analytical fit over $v$), and use those to test the resummation idea at 1SF.

\section{Numerical Method}
\label{sec:NumMethod}

Our numerical method is a hybrid scheme that combines frequency-domain (FD) data produced using the code of \cite{Whittall:2023xjp} with time-domain (TD) data produced using the code of \cite{Barack:2022pde}, to the effect of optimising the scattering angle calculation for speed and accuracy. Full details and performance analysis of the hybridization technique will be provided in a forthcoming paper \cite{Long_Whittall_inprep}. Here we will briefly review our TD and FD codes and the method of hybridization. 

\subsection{Time-domain code} \label{sec:TD}

Our TD method, developed in \cite{Barack:2022pde}, is based on characteristic evolution of Eq.\ (\ref{eq: KG with source}) in 1+1 dimensions. The scalar field $\Phi$ is first decomposed into angular spherical-harmonic modes, $\Phi=\sum_{\ell m}\phi_{\ell m}(t,r)Y_{\ell m}(\theta,\varphi)$. Each of the time-radial modal fields $\phi_{\ell m}(t,r)$ obeys a simple wave equation in 1+1 dimensions, sourced by a delta function supported on the geodesic path of the particle, which we choose in advance with fixed values of $(v, b)$. The initial-value evolution problem for each of the modal fields is solved numerically on a fixed grid based on null Eddington-Finkelstein coordinates using a second-order finite-difference scheme (detailed in Appendix B of Ref.~\cite{Barack:2022pde}). The evolution starts with characteristic initial data set to zero, and the data is recorded after the spurious (``junk") initial radiation has died away. The values of the modal fields and their time and radial derivatives are extracted along the geodesic worldline, and from them we compute the total $\ell$-mode derivatives $\nabla_\alpha \Phi_\ell  = \sum_{m=-\ell}^\ell \nabla_\alpha \phi_{\ell m}(t,r)Y_{\ell m}(\theta,\varphi)$ (evaluated along the worldline). 
The derivatives of the Detweiler-Whiting regular field are then constructed using standard mode-sum regularisation \cite{Barack:1999wf}:
\begin{align}
\nabla_\alpha \Phi^{\rm R} = \sum_{\ell=0}^\infty \Big[\nabla_\alpha \Phi_\ell - A_\alpha (\ell+\tfrac{1}{2}) - B_\alpha \Big]. \label{eq:mode_sum_formula}
\end{align}
Here $A_\alpha$ and $B_\alpha$ are the ``regularization parameters'', given analytically as functions along the worldline \cite{Barack:2001gx}. In practice we, of course, truncate the sum at a finite value, typically $\ell_{\rm max}=15$ in our TD code. The partial sum then has an error of $O(1/\ell_{\rm max})$. To improve the convergence of the mode sum, and thereby reduce this truncation error, we incorporate so-called ``high-order regularization parameters'', which successively remove higher-order terms in the $1/\ell$ expansion of $\nabla_\alpha \Phi_\ell$ \cite{Heffernan:2012su, Heffernan:2012vj}. We use this technique to reduce the partial-sum truncation error to mere $O(\ell_{\rm max}^{-7})$. Finally, given the derivatives of the regular field, we construct the SF along the scattering geodesic using Eq.\ (\ref{eq:cov_evolution}).

A typical TD run uses a numerical grid split up into cells of size $M/128\times M/128$, and produces clean SF data for $r_{\rm min}\leq r_p\leq r_{\rm fin}$ (on both legs of the orbits), where $r_{\min}$ is the periastron distance, and in our implementation $r_{\rm fin}$ ranges between $450M$ and $1250M$. The performance of the code degrades rapidly with increasing $\ell$ (due to resolution demands), and it is computationally prohibitive to go much beyond $\ell_{\rm max}=15$. Even so, we find that $\ell$-mode truncation error is usually subdominant in our code for $v$ values that are not too high, $v\lesssim 0.5$. At higher velocities, $\ell$-mode truncation becomes a limiting factor, as discussed below.  

Since the runtime scales like $r^2_{\rm fin}$, it is also prohibitive to increase $r_{\rm fin}$ much beyond the values stated above. As data is required for all $r\geq r_{\rm min}$, we fit the available SF data to a large-$r$ model of the form $F^{\rm self}_\alpha = c_{3}/r_p^{3} + c_{4}/r_p^{4}+...$, and use that to extrapolate the numerical results to $r_p\rightarrow\infty$. We fix the $c_i$ coefficients by fitting to the outer-most $10\%-20\%$ of the large-$r_p$ data. Varying the polynomial order and range of the fitting allows us to estimate the error of the fit. This is typically our dominant source of error in the scattering angle at small and medium $v$. 

To calculate the scattering angle we use the integral in Eq. (\ref{eq:1SF_scatter_angle}), recast as an integral in $r$ on the ingoing and outgoing legs. For the integrand we use the numerical data for $r_{\min} \leq r \leq r_{\rm fin}$ and the analytic fit for $r_{\rm fin} < r < \infty$, and perform the integration using Mathematica's default \texttt{NIntegrate} function, which suffices for our purposes.

\subsection{Frequency-domain code}\label{sec:FD}

Our FD code was developed in Ref.~\cite{Whittall:2023xjp}.
The modal fields $\phi_{\ell m}(t,r)$ are additionally decomposed into Fourier time-harmonics $\propto\! e^{-i\omega t}$, reducing the distributionally sourced partial differential equation to an ordinary differential equation with a function source. Reconstructing $\phi_{\ell m}(t,r)$ using solutions to this inhomogeneous equation is a-priori problematic, due to Gibbs ringing caused by the $\delta$-function source in the TD equation. A remedy is provided by applying the so-called method of extended homogeneous solutions (EHS), first introduced in Ref.~\cite{Barack:2008ms}. In this method, the TD fields $\phi_{\ell m}(t,r)$ and their derivatives are constructed along the particle's orbit from a sum of certain nonphysical homogeneous frequency modes $\phi_{\ell m\omega}(r)$, which is spectrally convergent. In the scattering problem, the method can be used to efficiently construct $\phi_{\ell m}(t,r)$ and its derivatives in the region $r \leq r_p(t)$, sufficient for a SF calculation via mode-sum regularization.

The calculation of $\phi_{\ell m\omega}(r)$ entails the evaluation of certain normalization factors $C_{\ell m\omega}$, which are expressed as integrals over the (infinite) radial extent of the orbit. In our FD code these integrals are truncated at a radius $r_{\rm max} = 2000M$, and the truncation error is reduced by using four successive integration by parts to increase the decay rate of the integrand to $O(r^{-5})$ as $r\rightarrow \infty$. For $v \geq 0.3$, we additionally add an analytic approximation to the neglected $r > r_{\rm max}$ portion of the integral. The values of the normalization integrals are stored at discrete frequencies with spacing $M\Delta\omega = 1.25 \times 10^{-3}$, and intermediate frequencies are calculated as needed using interpolation. The inverse Fourier integrals are numerically evaluated to reconstruct the derivatives of $\phi_{\ell m}(t,r)$ along the orbit, and the SF is then calculated using the mode-sum formula (\ref{eq:mode_sum_formula}) and Eq.\ (\ref{eq:cov_evolution}).

Comparisons with TD results suggest that the FD code is highly accurate in the strong-field, near-periapsis portion of the orbit, significantly outperforming the TD code in this region (see, e.g., Fig.\ 9 of \cite{Whittall:2023xjp}). A particular advantage of the FD code is its ability to access high $\ell$-modes. For this project we have reliably reached $\ell = 25$, at least at radii not too large.  

As the particle moves outwards along its orbit, however, the large-$\ell$ modes calculated using the FD code begin to rapidly lose accuracy, with progressively smaller values of $\ell$ becoming affected with increasing radius. As discussed in Sec.~IX of Ref.~\cite{Whittall:2023xjp}, this is a numerical issue resulting from increasing cancellation between low-frequency modes of the EHS with increasing $r$, a phenomenon that was first reported in studies of the gravitational SF along bound geodesics in the Kerr spacetime \cite{vandeMeent:2017bcc}. In mitigation, we dynamically truncate the $\ell$-mode sum where cancellation-induced error is deemed too great. The original algorithm to achieve this is outlined in Sec. VII C of Ref.~\cite{Whittall:2023xjp}. According to this procedure, at a given orbital location, we first calculate all $\ell$ modes with $\ell \leq \ell_{\rm min}$, where $\ell_{\rm min}$ is some minimum number of modes to be included, taken as 5 in the present work. Successive $\ell$ modes are then calculated one by one, and added to the mode sum so long as the absolute value of the regularized contribution to the SF is {\em decreasing}. Additional clauses exist to identify and include $\ell$-modes where a transient increase in magnitude follows a legitimate change of sign. The mode sum is truncated when the algorithm excludes an $\ell$-mode for the first time, or when it reaches $\ell=25$. The progressively early truncation of the mode sum at larger radii evades the cancellation problem, but causes growing errors in the SF calculation with increasing $r$, eventually causing the accuracy to fall below that of the TD code. 

In standard SF calculations for bound orbits (or for scattering orbits at low velocity), the $\ell$-mode contributions to the SF fall off with a power law $\ell^{-k}$, where $k>0$ depends on details of the regularization procedure applied. In this work we achieve $k=8$ by subtracting all analytically known regularization parameters in Eq.~\eqref{eq:mode_sum_formula}, as mentioned above. Fundamentally, the power-law distribution reflects the structure of the field singularity at the particle.  In producing FD data for this project, we have found an interesting new structure at large $\ell$, only manifest at high velocity, which may be attributed to a beaming effect. The new feature, which we now briefly discuss, requires extra care in handling large-$\ell$ contributions at high velocity. 

\begin{figure}[htb]
\centering
\includegraphics[width=\linewidth]{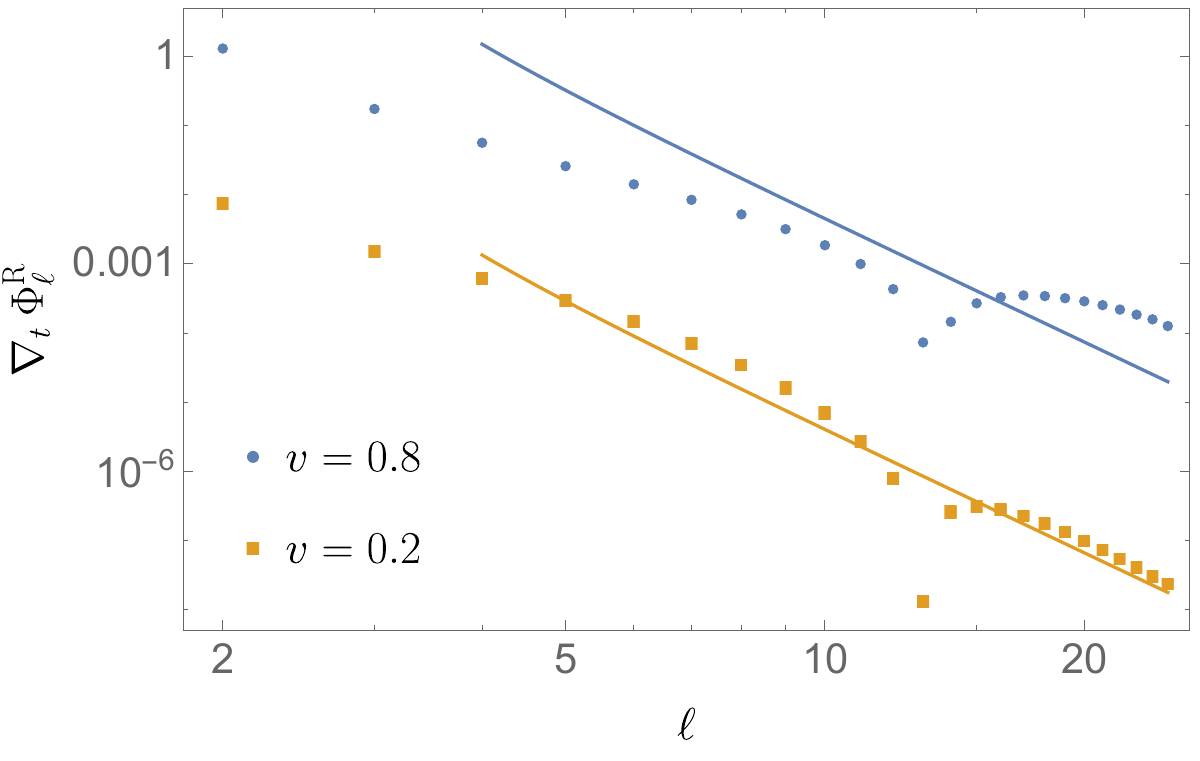}
    \caption{Regularized $\ell$-mode contributions to $\nabla_t \Phi^R$ at positions $r_p = 4.7M$ and $r_p = 4.2M$ along the outbound legs of the orbits with parameters $(v,b) = (0.8, 6.07387M)$ and $(0.2, 20.3825M)$, respectively (both orbits have $\delta b \approx 0.0005M$). Also plotted are the analytical predictions for the large-$\ell$ asymptotics of these contributions. The dips near $\ell=13$ correspond to a sign change in both datasets. At high velocity, the $\ell$-mode contributions take longer to settle down to the predicted asymptotic $\ell^{-6}$ decay, with distinct large-$\ell$ peaks visible at intermediate radii. Such large-$\ell$ spectral ``bumps'', which may be a signature of radiation beaming, complicate computations at large $v$.}
\label{Fig:ellBump}
\end{figure}

Figure ~\ref{Fig:ellBump} provides an illustration. It shows the value of the mode-sum summand as a function of $\ell$, in the example of $\alpha=t$ at a certain point along a certain near-separatrix scattering orbit with $v = 0.8$. For comparison, the same is shown for a lower-velocity orbit with $v=0.2$. For the purposes of this illustration we have subtracted fewer of the higher order regularization parameters, such that the terms in the mode-sum are expected to fall only as $\ell^{-6}$. Doing so enables us to use the un-subtracted regularization term as an analytical prediction for the asymptotic behavior of the terms in the mode sum. From the figure we can see that in the $v = 0.2$ case the contributions approach the asymptotic prediction closely after $\ell \approx 15$. For $v = 0.8$, however, the magnitude of modal contributions picks up again at around $\ell=14$ to form a broad ``bump'' in the angular spectrum. The ultimate $\ell^{-6}$ tail presumably develops only at greater values of $\ell$, beyond the range accessible to us here.

The potential occurrence of such late ``bumps'' in the angular spectrum somewhat complicates our automatic $\ell$-mode truncation algorithm.  We have inserted a set of clauses that legitimize the inclusion of $\ell$-modes associated with these bumps. The ability to include the contributions from larger values of $\ell$ gives the FD code a significant advantage over the TD code at large $v$. 

\subsection{Hybridization of TD and FD results}\label{sec:hybrid_method}
The broadness of the $\ell$-mode power distribution at large velocity, illustrated in Fig.~\ref{Fig:ellBump}, has important practical implications. When truncating the mode sum at $\ell_{\rm max} = 15$ (the practical limit of the TD code), the error compared to $\ell_{\rm max} = 25$ is observed to be on the order of several percent in some instances. This is significantly larger than other estimated numerical errors in our SF calculations. Fortunately, the problem is greatest in the immediate vicinity of the periastron, precisely where the FD method has high-precision access to large-$\ell$ modes. Conversely, the problem becomes less significant at large radii, where the TD code outperforms the FD code. This naturally suggests an optimization approach that uses the appropriate sets of data in each regime.

For a given geodesic orbit, the TD-FD data hybridization is performed in the following way. First, the TD and FD codes are run separately. The SF is extracted from the TD code at radii $\rmin \leq r_p \leq r_{\rm fin}$, and it is also calculated at a grid of radii in $\rmin \leq r_p \leq 50M$ using the FD code. The output of the FD code includes the truncation value $\ell_{\rm max}$ used at each position, and from this we identify the largest radius $r_{\rm switch}$ such that $\ell_{\rm max} \geq 15$ for all components of the force at all radii $r_p \leq r_{\rm switch}$. For the orbits tested in this article, we always find $r_{\rm switch} < 50M$, justifying the appropriateness of our FD radial truncation. The scatter angle is then calculated by recasting Eq.~\eqref{eq:1SF_scatter_angle} as an integral over radius $r$, and performing the sections over $\rmin \leq r \leq r_{\rm switch}$ and $r_{\rm switch} < r < r_{\rm fin}$ separately using SF data from the FD and TD codes, respectively. The $r > r_{\rm fin}$ portion is again approximated by fitting the SF to the final 10-20\% of the TD data.

\begin{figure}[htb]
\centering
\includegraphics[width=\linewidth]{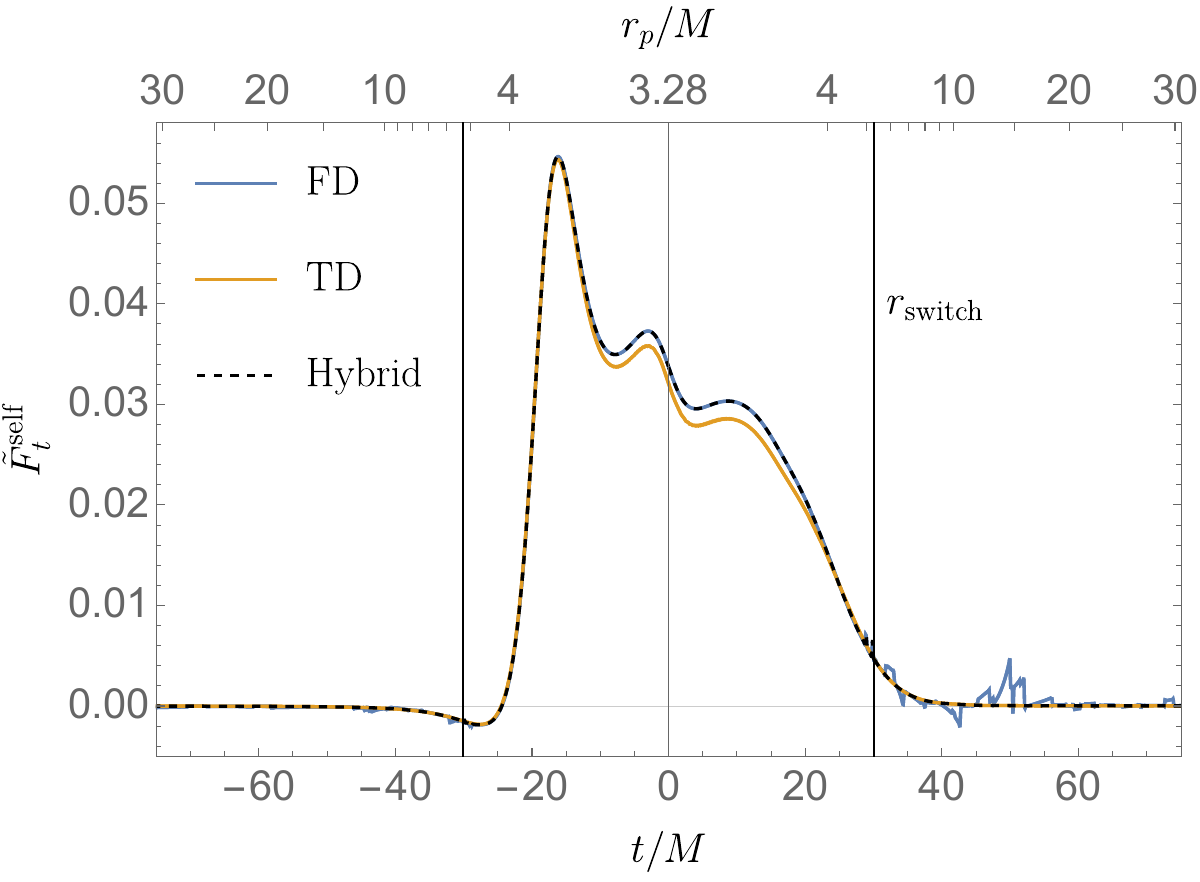}
\caption{The time component of the SF for the orbit $(v,b)=(0.7,6.71307M)$ (with $\delta b \approx 0.001M$), as calculated with the FD (blue) and TD (orange) codes. The visible difference between the two codes near periastron is due to the omission in the TD calculation of beaming features at $\ell>15$, beyond the reach of our TD code. The hybrid model is shown in the black dashed line, with $r_{\rm switch}\approx 5.28M$ marked by the vertical black lines. The model uses the TD data for $r>r_{\rm switch}$ and the FD data for $r<r_{\rm switch}$. The noisy features visible in the FD data are artifacts of our dynamical mode-sum truncation algorithm; they occur mostly at  $r>r_{\rm switch}$ where they do not impact our hybrid model.}
\label{Fig:Hybrid}
\end{figure}

Fig.\ \ref{Fig:Hybrid} illustrates the hybridization of TD and FD data for a sample scattering orbit with $v=0.7$.  The agreement is visibly not good near the periastron, where the TD data fails to account for large-$\ell$ beamed power. Indicated in the figure is the radius $r_{\rm switch}$ where we switch from FD data ($r<r_{\rm switch}$) to TD data ($r>r_{\rm switch}$).

\section{Numerical calculation of $A_1(v)$}
\label{sec:ResultsA1}

Equations \eqref{A1_diss} and \eqref{A1_cons} prescribe the direct calculation of $A_1^{\rm diss}(v)$ and $A_1^{\rm cons}(v)$, requiring as input only the SF along the critical orbit $b = b_c(v)$. Unfortunately, the existing TD and FD codes are currently configured to handle only non-critical orbits, with $b > b_c(v)$. Extending to the critical orbit requires non-trivial modifications to both codes. Instead, we opted here to calculate the values of $A_1(v)$ indirectly by extrapolating from a sequence of geodesics $b \rightarrow b_c(v)^+$ with fixed $v$. We will return to the question of directly calculating the SF along critical orbits in Sec.~\ref{sec:Conc}.

We fixed a grid of velocities in the range $0.15 \leq v \leq 0.7$ with spacing $\Delta v = 0.05$. At lower velocities, the transition to the asymptotic behavior $\chi\sim 1/\delta b$ is delayed until smaller values of $\delta b$, complicating the extrapolation. At higher velocities, it takes longer for the initial junk radiation to separate from the particle in the TD simulation, forcing us to start at a larger initial radius and hence increasing computational cost. For each velocity included in our sample, we ran the TD and FD codes to calculate the SF along each of the orbits with $b = b_c(v) + \delta b$, where $\delta b$ takes values in $\{0.0005,\allowbreak 0.001,\allowbreak 0.0022,\allowbreak 0.005,\allowbreak 0.01,\allowbreak 0.022,\allowbreak 0.05,\allowbreak 0.1,\allowbreak 0.22,\allowbreak 0.5,\allowbreak 1\}$. The values of $\chi^{\rm cons}$ and $\chi^{\rm diss}$ were then calculated separately for each orbit using the hybrid method outlined in Sec.~\ref{sec:hybrid_method}. 

\begin{figure}[htb]
\centering
\includegraphics[width=\linewidth]{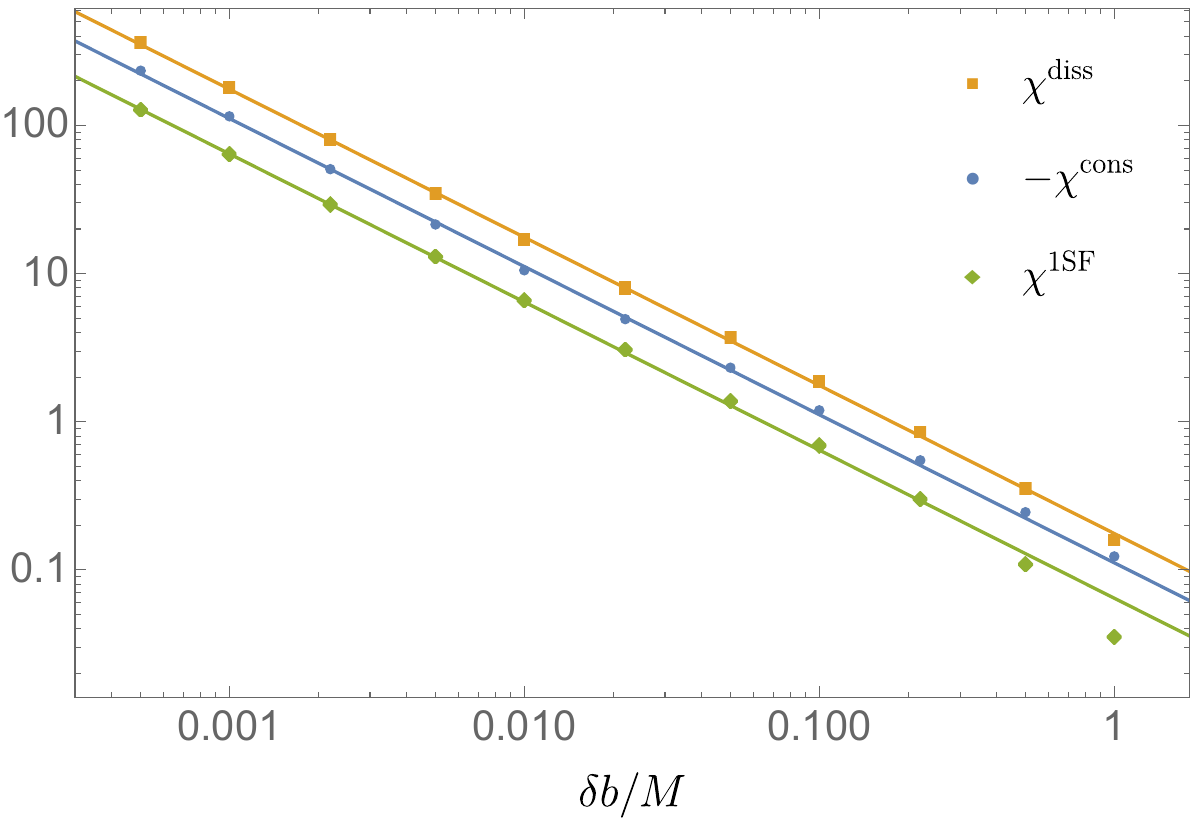}
\caption{Strong-field numerical results for the dissipative (orange), conservative (blue), and total (green) pieces of the scattering angle for an initial velocity $v=0.5$. The values were generated with the hybrid model, with error bars too small to be discerned on the scale of this plot. The solid lines are the extracted divergences of the form $\sim A_1 (b_c/\delta b)$. The fitted values of $A_1$ are given in Table \ref{tab:A1_vs_v}.}
\label{Fig:extrap_plot}
\end{figure}

Figure~\ref{Fig:extrap_plot} displays $\chi^{\rm 1SF}$, $\chi^{\rm diss}$ and $\chi^{\rm cons}$, plotted as functions of $\delta b$ at fixed $v = 0.5$, illustrating the $1/\delta b$ divergence.  For each value of $v$ in our sample we fit the numerical dataset to the expression on the right hand side of Eq.~\eqref{chi1SF_asy}, to obtain an estimate of $A_1(v)$. For this purpose we use Mathematica's \texttt{NonlinearModelFit} function, weighting each data point by $1/\epsilon_{\rm tail}^2$, where $\epsilon_{\rm tail}$ is the estimated error in the scattering angle due to the analytic fit to the SF at large radius. This routine returns an estimate for the value of $A_1(v)$, together with an estimate of the fitting error in this value. We perform the fit for the conservative and dissipative pieces separately, and then calculate $A_1(v) = A_1^{\rm diss}(v) + A_1^{\rm cons}(v)$. This approach is potentially more accurate, because, as illustrated in Fig.~\ref{Fig:extrap_plot}, the opposite signs of the conservative and dissipative contributions cause the total scattering angle to approach the asymptotic $1/\delta b$ behavior somewhat more slowly than for $\chi^{\rm diss}$ and $\chi^{\rm cons}$ individually.

Our fitting procedure is as follows. For each velocity, values of $A_1^{\rm diss}$ and $A_1^{\rm cons}$ are obtained by fitting to the $N$ smallest values of $\delta b$ in our sample, for $N = 3, ..., N_{\rm max}$. We took $N_{\rm max}=8$ for $v \geq 0.3$, and $N_{\rm max}=6$ for $v < 0.3$ where the $\sim 1/\delta b$ trend is observed to break down at lowers values of $\delta b$. A best estimate for each velocity was obtained by fitting a constant value to these individual fits, again using \texttt{NonlinearModelFit}, weighting the individual fits by the inverse squares of their estimated errors. The final error bar on $A_1$ (for each $v$) was conservatively estimated as the range between the largest and smallest values in our sample of individual fits, including their individual error bars.

\begin{figure}[htb]
\centering
\includegraphics[width=\linewidth]{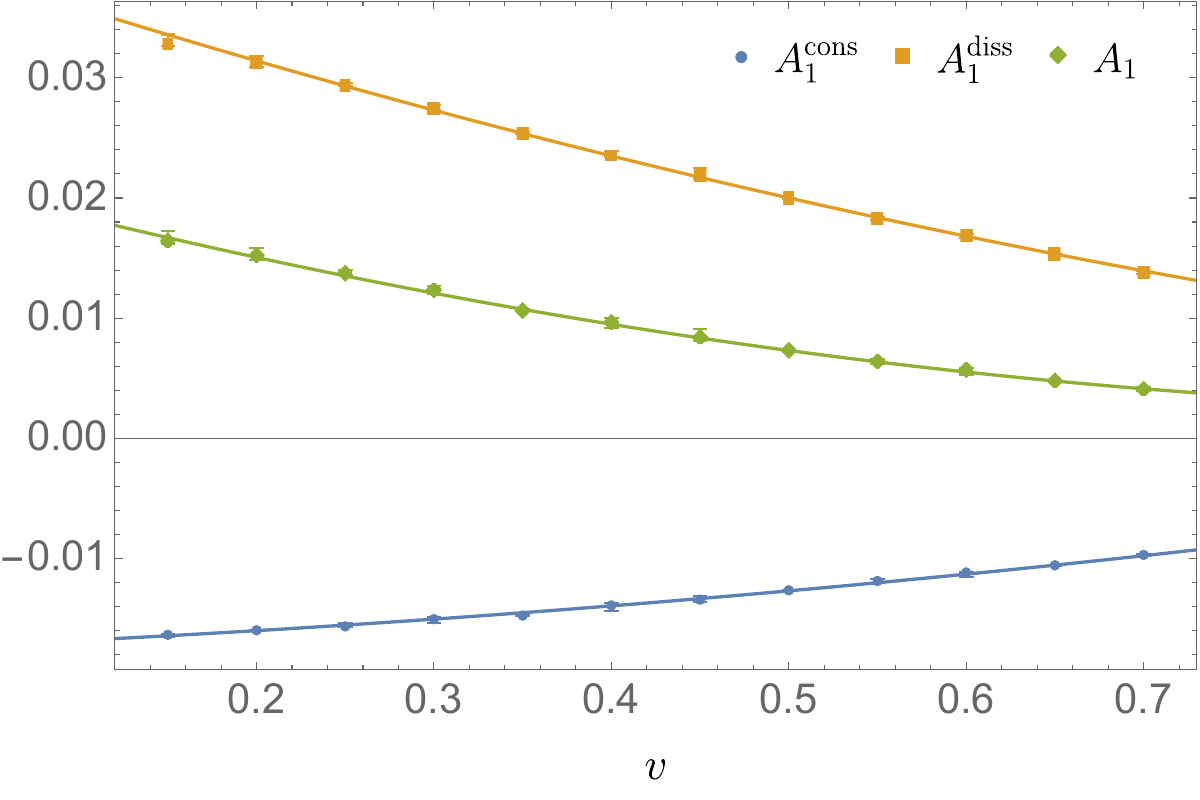}
\caption{Values of the 1SF divergence parameter $A_1(v)$ for the conservative (blue), dissipative (orange), and total (green) pieces as a function of initial velocity $v$. The error bars on the values are shown but are barely visible on this scale. Shown in solid lines are the best-fit curves given in Eqs.~\eqref{eq:A1fit_total}--\eqref{eq:A1fit_diss}.}
\label{Fig:A1_vs_v}
\end{figure}

\begin{table}
\begin{ruledtabular}
\begin{tabular}{l l l l}
 $v$ & $A_1$ & $A_1^{\rm cons}$ & $A_1^{\rm diss}$\\
 \hline\vspace{-2mm}\\
 $0.15$ & $0.01642_{-22}^{+80}$ & $-0.01637_{-14}^{+04}$ & $0.03280_{-17}^{+80}$ \\[3pt]
 $0.20$ & $0.01522_{-35}^{+60}$ & $-0.01600_{-6}^{+4}$ & $0.03122_{-34}^{+60}$ \\[3pt]
 $0.25$ & $0.01373_{-11}^{+31}$ & $-0.01568_{-05}^{+24}$ & $0.02941_{-10}^{+19}$ \\[3pt]
 $0.30$ & $0.01234_{-31}^{+40}$ & $-0.01507_{-29}^{+16}$ & $0.02741_{-10}^{+33}$ \\[3pt]
 $0.35$ & $0.01062_{-07}^{+12}$ & $-0.01477_{-4}^{+8}$ & $0.02539_{-5}^{+9}$ \\[3pt]
 $0.40$ & $0.0096_{-5}^{+4}$ & $-0.01393_{-50}^{+17}$ & $0.02357_{-19}^{+32}$ \\[3pt]
 $0.45$ & $0.00839_{-32}^{+70}$ & $-0.01345_{-21}^{+29}$ & $0.02184_{-25}^{+70}$ \\[3pt]
 $0.50$ & $0.00731_{-4}^{+5}$ & $-0.012670_{-27}^{+26}$ & $0.019978_{-35}^{+50}$ \\[3pt]
 $0.55$ & $0.00639_{-25}^{+21}$ & $-0.01190_{-20}^{+14}$ & $0.01829_{-15}^{+16}$ \\[3pt]
 $0.60$ & $0.00570_{-40}^{+15}$ & $-0.01119_{-35}^{+03}$ & $0.01689_{-20}^{+15}$ \\[3pt]
 $0.65$ & $0.004761_{-26}^{+26}$ & $-0.010585_{-18}^{+18}$ & $0.015345_{-20}^{+19}$ \\[3pt]
 $0.70$ & $0.00407_{-15}^{+15}$ & $-0.00972_{-11}^{+11}$ & $0.01379_{-10}^{+11}$ \\[3pt]
\end{tabular} 
\end{ruledtabular} 
\caption{Calculated values of $A_1(v)$, $A_1^{\rm cons}(v)$ and $A_1^{\rm diss}(v)$, with estimated error bars on the last displayed decimals (e.g., $0.01642_{-22}^{+80}$ means $0.01642_{-0.00022}^{+0.00080}$). The error bars for $A_1(v)$ are obtained by adding the error bars of the conservative and dissipative pieces in quadrature.}
\label{tab:A1_vs_v}
\end{table}

The resulting values of $A_1(v)$, with error bars, are displayed in Fig.~\ref{Fig:A1_vs_v} and tabulated in Table~\ref{tab:A1_vs_v}. Also included in Fig.~\ref{Fig:A1_vs_v} are the best-fit curves
\begin{align}
    A_1(v) &\approx 0.0222 - 0.0398v + 0.0199v^2,\label{eq:A1fit_total}\\
    A_1^{\rm cons}(v) &\approx -0.0175 + 0.0060v + 0.0072v^2, \label{eq:A1fit_cons}\\
    A_1^{\rm diss}(v) &\approx 0.0406 - 0.0488v + 0.0154v^2, \label{eq:A1fit_diss}
\end{align}
obtained by fitting functions of the form $a+bv+cv^2$ to the numerical data, weighting each point by the inverse square of the size of its error bar.

\section{Resummation results}
\label{sec:ResultsResum}

We now test the performance of our resummation formula (\ref{eq:ResumFormula2}), with the $A_1$ values obtained above, using the numerical SF scattering angle data as benchmark. 

\begin{figure}
\begin{center}
\includegraphics[width=\linewidth]{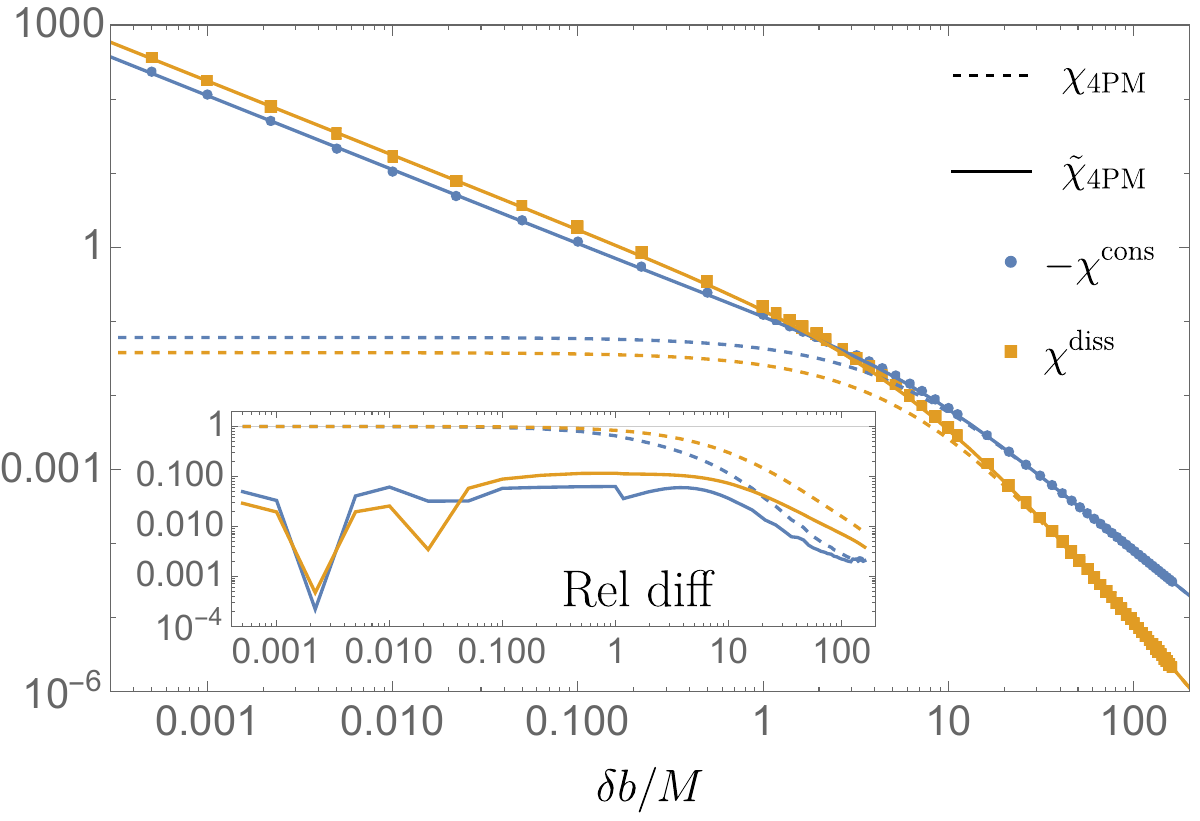} 
\caption{Comparisons of plain (dashed) and resummed (solid) 4PM expressions for the conservative (blue) and dissipative (orange) contributions to the scalar-field SF correction to the scattering angle at $v=0.5$. The inset shows the relative difference between the PM results and the ``exact" numerical values. The resummation formula, given in Eq.\ (\ref{eq:ResumFormula2}), is used with the values of the SF singularity coefficients $A_1^{\rm cons}$ and $A_1^{\rm diss}$ given in Table \ref{tab:A1_vs_v}. The numerical values are calculated using the hybrid model for $\delta b/M\leq 1$ and the TD model for larger impact parameters. Numerical errors are too small to be resolved on the scale of the main plot. }
\label{fig:ResumPM_vs_SF_v0.5}
\end{center}
\end{figure}

Figure \ref{fig:ResumPM_vs_SF_v0.5} shows numerical $\chi^{\rm cons}$ and $\chi^{\rm diss}$ values as functions of $b$ at fixed $v=0.5$, along with the plain and resummed PM expressions. As previously demonstrated, the plain PM formula matches the ``exact" numerical values to within a few percent in the weak field, but the accuracy quickly degrades when moving towards the strong field. However, the resummed PM expressions are uniformly accurate across the entire domain. Notably, the resummation generally appears to improve the performance of the PM formulas even in the weak-field regime. Similar results are obtained for all other values of $v$ sampled in our work. 

\begin{figure}
\begin{center}
\includegraphics[width=\linewidth]{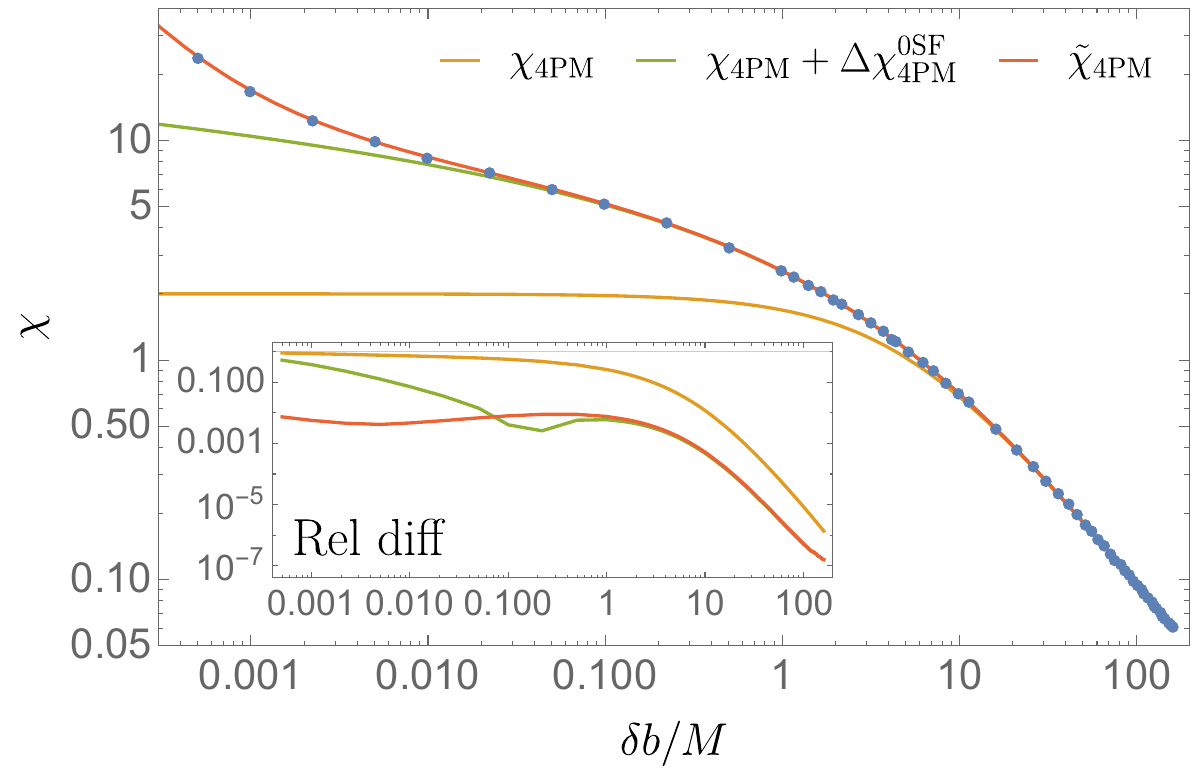} 
\caption{The total scattering angle (geodesic + SF correction) for $\epsilon=0.1$ and $v=0.5$. We compare here numerical values obtained using our SF calculation (blue dots) to corresponding PM expressions with no resummation (orange), geodesic resummation only (green) and the full 1SF resummation (red). The relative difference between the results from each PM expression and the numerical values are shown in the inset, interpolated. The hybrid model was used to calculate the angles with $\delta b/M\leq 1$ and the TD model was used for all other values. Numerical errors are too small to be resolved in the main plot. 
}  
\label{v0.2_Resum}
\end{center}
\end{figure}

As a final illustration, let us add together the geodesic and 1SF contributions, to show the combined effect of resumming both the logarithmic (geodesic) and power-law (SF-induced) divergences. Figure \ref{v0.2_Resum} shows the total angle $\chi = \chi^{\rm 0SF} + \epsilon\,\chi^{\rm 1SF}$ for fixed $\epsilon=0.1$ and $v=0.5$, along with the analytic results with no resummation (orange), only geodesic resummation (green) and the full 1SF resummation (red). Both resummations appear to increase the accuracy in the weak field by approximately an order of magnitude relative to the base PM expressions. Differences between the resummations become manifest in the strong-field regime where the $1/\delta b$ divergence starts to dominate. The full resummation captures the scattering angle with at least $\sim 1\%$ precision, including in regions where the plain PM expansion completely breaks down. 

\section{Conclusion}
\label{sec:Conc}

We have presented and tested here a technique for improving the faithfulness of PM expressions for the scattering angle in the strong-field regime. The dramatic improvement achieved can be appreciated from a comparison of Fig.\ \ref{Fig:PM_vs_SF} (raw PM expansion) with Figs.\ \ref{fig:ResumPM_vs_SF_v0.5} and  \ref{v0.2_Resum} (resummed PM expansion). Notably, in the examples considered we have found that our resummation procedure improves the faithfulness of the PM expressions (at given PM order) uniformly across the parameter space, and even in the weak-field regime. The procedure should offer a computationally cheap way of producing a highly accurate semi-analytical model of black hole scattering dynamics, at least in the small mass-ratio regime. As an aside, our work provides further illustration of the utility of SF calculations in benchmarking strong-field aspects of the two-body dynamics. 

The particular strong-field aspect utilized here is the leading SF correction to the divergent behavior of the scattering angle at the capture threshold, encapsulated in the coefficient $A_1(v)$. This function can be obtained as prescribed in Eq.~(\ref{A1}), by integrating a certain combination of SF components along critical geodesics. In our numerical demonstration we have not directly integrated along critical orbits, but instead chose to approach the critical limit along a sequence of scattering geodesics (for each fixed value of $v$ in our sample). We did so in order to be able to use our existing TD and FD codes with minimal adaptation, at the expense of having to produce $\sim\! 10$ times the amount of SF data that would be required integrating directly along critical geodesics. A direct implementation of Eq.~(\ref{A1}) would be more economical as well as more precise, and should be considered for future work (e.g., when ultimately implementing the resummation idea with the gravitational SF for a black hole binary). This would require the development of customized versions of our codes that can deal with the special nature of critical geodesics, each consisting of two disjoint segments that asymptote to an unstable circular orbit.  In the TD framework, the evolution code would need to be run twice for each critical geodesic, with an appropriate truncation as the geodesic settles into (emerges from) the asymptotic circular motion. This can be based on the method of Ref.\ \cite{Barack:2019agd}, where a similar scenario was dealt with in the gravitational problem with $v=0$. Critical orbits have not yet been considered in FD self-force calculations. Here the task would be to correctly account for the special spectral features of the perturbation field, which involve a superposed delta-function component from the asymptotic circular whirl. This is yet to be formulated and attempted computationally. 

Our numerical method in this work does incorporate several new developments, primarily the introduction of our hybrid TD-FD scheme. The idea is to combine TD and FD self-force data along each individual scattering orbit to the effect of optimizing the calculation for accuracy, through a judicious consideration of each method's different performance profile as a function of radius and mode number. Hybridization enabled us to achieve a greater precision over a greater portion of the parameter space than would be achievable with either the TD or FD codes alone. The hybridization method should have a merit beyond just the scope of this project, and we intend to continue its development. A forthcoming paper \cite{Long_Whittall_inprep} will provide a detailed analysis of the method and its performance across the scattering parameter space.  

Ultimately, of course, the goal is to apply our methods to the binary black hole problem. We are examining several alternative avenues. One is based on an extension of the (Lorenz-gauge) TD code of Ref.\ \cite{Barack:2019agd} from the special case of the critical orbit with $v=0$ to general scattering orbits. Another may involve an FD variant of the same code, which should allow improved computational precision. An alternative approach is based on metric reconstruction from curvature scalars \cite{Long:2021ufh}, which, in principle, may be implemented in either the time or frequency domains. Work is in progress to develop an efficient, modern framework for scattering-orbit calculations based on a TD Teukolsky solver with hyperboloidal slicing and compactification \cite{Macedo_etal_inprep}.

\section*{Acknowledgements}

We are grateful to Maarten van de Meent for conversations that inspired the method of this work. We thank Davide Usseglio, as well as Zvi Bern,  Enrico Herrmann, Julio Parra-Martinez, Radu Roiban, Michael S. Ruf and Chia-Hsien Shen for many useful discussions.  CW acknowledges support from EPSRC through Grant No. EP/V520056/1. We acknowledge the use of the IRIDIS High Performance Computing Facility, and associated support services at the University of Southampton, in the completion of this work. This work makes use of the Black Hole Perturbation Toolkit \cite{BHPToolkit}.

\appendix 

\section{Logarithmic divergence of $\chi^{\rm 0SF}$}\label{App}

In this appendix we show the derivation of Eq.\ (\ref{chi0_bv}) for the logarithmic divergence of the scattering angle near the separatrix in the geodesic limit. 

Starting with the expression (\ref{eq:geodesic_scatter_angle}) for $\chi^{\rm 0SF}$ in terms of an elliptic function, we expand in $p$ about $p_c(e)=6+2e$ (at fixed $e$), to find
\begin{equation}\label{chi0_ep}
\chi^{\rm 0SF} \simeq -\left(\frac{6+2e}{e}\right)^{1/2}\log\left(\delta p/p_c\right) + {\rm const},
\end{equation}
where $\delta p:= p-p_c(e)$. We need now to (i) express the $e$-dependent coefficient in terms of $v$ on the separatrix, and (ii) express $\delta p/p_c$ (fixed $e$) in terms of $\delta b/b_c$ (fixed $v$) inside the logarithm. For (i), we use Eq.\ (\ref{eq:ELepRelation}) to write
\begin{equation}\label{Lofe}
L_c(e)/M= \frac{p_c(e)}{\sqrt{p_c(e)-3-e^2}} = \frac{6+2e}{\sqrt{(3-e)(1+e)}}; 
\end{equation}
then invert to obtain $e$ in terms of $L_c$ on the separatrix. This yields
\begin{align}
-\left(\frac{6+2e}{e}\right)^{1/2} = 
& -2\left(1-\frac{12M^2}{L_c^2}\right)^{-1/4} \nonumber\\
=& -2\left(1-\frac{12M^2(1-v^2)}{v^2 b_c^2}\right)^{-1/4},
\end{align}
where in the second equality we have used Eqs.\ (\ref{v}) and (\ref{eq:bdef}).
Recall that $1<e<3$ for scattering geodesics, so expressions like those in (\ref{Lofe}) make sense.

To achieve goal (ii) above, we write 
\begin{equation}
\delta b = b-b_c(v) = \frac{L}{\sqrt{E^2-1}} - b_c(v(E)),
\end{equation}
again using Eqs.\ (\ref{v}) and (\ref{eq:bdef}), then 
substitute for $b_c(v(E))$ from Eq.\ (\ref{bc}), and finally for $E$ and $L$ in terms of $p$ and $e$ from Eq.\ (\ref{eq:ELepRelation}). The resulting function of $p$ and $e$ we expand in $p$ about $p_c(e)=6+2e$ at fixed $e$. We find
\begin{equation}\label{deltab_deltap}
\delta b = \frac{e (3+e)^{3/2}M}{16 (1+e)^3 (e-1)^{1/2}}(\delta p)^2 + O(\delta p^3),
\end{equation}
noting the linear term vanishes.
Hence 
\begin{equation}
\log(\delta p/p_c) = \frac{1}{2}\log(\delta b/b_c) + {\rm const},
\end{equation}
and Eq.\ (\ref{chi0_ep}) produces (\ref{chi0_bv}), with the coefficient $A_0(v)$ given in Eq.\ (\ref{A0}).

\bibliography{biblio}

\end{document}